\documentclass{article}

\usepackage{microtype}
\usepackage{graphicx}
\usepackage{subfigure}
\usepackage{multirow}
\usepackage{booktabs} 
\usepackage{amsthm,amsmath,amssymb}
\usepackage{mathrsfs}
\usepackage{float} 
\usepackage{subfigure}
\usepackage{hyperref}

\usepackage[table,xcdraw]{xcolor}
\usepackage[normalem]{ulem}
\usepackage[accepted]{icml2020}
\icmltitlerunning{Submission and Formatting Instructions for ICML 2020}

\begin{document}

\twocolumn[
\icmltitle{ Heterogeneous Graph Neural Network for Recommendation}



\icmlsetsymbol{contact}{*}
\icmlsetsymbol{equal}{+}

\begin{icmlauthorlist}
\icmlauthor{Jinghan Shi}{equal}
\icmlauthor{Houye Ji}{equal}
\icmlauthor{Chuan Shi}{contact}
\icmlauthor{Xiao Wang}{}
\icmlauthor{Zhiqiang Zhang}{}
\icmlauthor{Jun Zhou}{}

\end{icmlauthorlist}


\icmlcorrespondingauthor{Jinghan Shi}{superali@bupt.edu.cn}
\icmlcorrespondingauthor{Houye Ji}{jhy1993@bupt.edu.cn}
\icmlcorrespondingauthor{Chuan Shi}{shichuan@bupt.edu.cn}
\icmlcorrespondingauthor{Xiao Wang}{xiaowang@bupt.edu.cn}
\icmlcorrespondingauthor{Zhiqiang Zhang}{zzqsmall@gmail.com}
\icmlcorrespondingauthor{Jun Zhou}{jun.zhoujun@antfin.com}

\icmlkeywords{Machine Learning, ICML}

\vskip 0.3in
]



\printAffiliationsAndNotice{\icmlEqualContribution} 

\begin{abstract}
The prosperous development of  e-commerce has spawned diverse recommendation  systems. As a matter of fact, there exist rich and complex interactions among various types of nodes in real-world recommendation systems, which can be  constructed   as  heterogeneous graphs.  How learn representative node embedding is the basis and core of the personalized recommendation system. Meta-path is a widely used structure to capture the semantics beneath such interactions and show potential  ability in improving node embedding.
In this paper, we  propose  \textbf{H}eterogeneous
\textbf{G}raph neural network for  \textbf{Rec}ommendation (HGRec) which injects  high-order semantic into node embedding  via aggregating multi-hops meta-path based neighbors and fuses rich semantics via multiple meta-paths based on attention mechanism to get comprehensive node embedding. 
Experimental results    demonstrate the importance of rich high-order semantics and also show the potentially
good interpretability of HGRec.
\end{abstract}

\section{Introduction}
In the era of information explosion, the recommender system has
become one of the most effective ways to help users to discover
what they are interested in enormous data. Generally speaking, the recommender systems usually follow   two steps: learn  vectorized representations (aka. embeddings) of users and items and then model interactions among them (e.g., whether a user buy an item).  Collaborative filtering (CF) learns  node embedding based on the historical interactions on user-item bipartite  graph and performs item recommendation based on the parameters.

As a matter of fact, there exist diverse relations among various types of nodes (e.g., buy relation and social relation) in real-world recommendation scenario, also widely known as heterogeneous graph \cite{17tkde_shi}.  Taking the dataset Movielens as an example,  it contains three types
of nodes include movie, user and genre.  
Meta-path \cite{11vldb_pathsim}, a composite relation connecting two objects, is a
widely used structure to capture the semantics. 
The semantics revealed by different meta-paths are able to describe the  characteristics of nodes from different aspects. For example, meta-path User-Movie (U-M)  describes the preference of user, while  meta-path User-User (U-U)  describes social influence among users. 
Besides basic meta-path, multi-hop meta-path (e.g., U-U-U) which captures high-order semantics and enrich the connections among users is able to improve the node embedding and alleviate the cold-start problem.

Based on the above analysis, when designing heterogeneous graph neural network for recommendation, we need to address the following  requirements.
\begin{itemize}
\item \textbf{Heterogeneity of graph.} The heterogeneity is an intrinsic property of heterogeneous graph, i.e., various types of nodes and edges.  How to handle such complex structural information for recommendation is an urgent problem that needs to be solved.
\item \textbf{High-order semantic preservation. } High-order semantic information which captures diverse  long-term dependencies among nodes  plays the key role in improving node embedding and alleviating the cold-start problem in recommender system. How to inject high-order semantic into node embedding is a fundamental problem in recommender system.
\item \textbf{Rich semantics fusion.}  Different meaningful and complex semantic information are involved in heterogeneous graph, which are usually reflected by diverse meta-paths. For example, meta-path U-M and U-U can describe the preference and social influence of user and then  comprehensively describe the characteristics of user  from different aspects.   How to select the most meaningful meta-paths and fuse rich semantics to improve node embedding is an open problem.
\end{itemize}
In this paper, we  propose  \textbf{H}eterogeneous
\textbf{G}raph neural network for  \textbf{Rec}ommendation, named HGRec, which mainly considers high-order semantic preservation and rich semantics fusion. Specifically, 
semantic aggregation layer injects high-order semantic  into node embedding via multi-hop meta-path and semantic fusion layer fuse rich semantics revealed by multiple meta-paths. 
After that, the overall model can be optimized via back propagation in an end-to-end manner.

The contributions of our work are summarized as follows:
\begin{itemize}
\item We  highlight the critical importance of rich high-order semantics in improving node embedding for recommendation system.
\item We propose a heterogeneous graph neural network based recommendation system, which explicitly injects high-order semantic into node embedding via multi-hops meta-path  and fuses rich semantics via multiple meta-paths for comprehensive node embedding.
\item  Empirical studies on real-world heterogeneous graphs 
 demonstrate the state-of-the-art performance of HGRec and potentially
good interpretability for the recommendation results.

\end{itemize}
\section{METHODOLOGY}
In this section, we present the proposed model \textbf{H}eterogeneous
\textbf{G}raph neural network for  \textbf{Rec}ommendation (HGRec). The basic idea of HGRec is to learn representative node
embedding of  users and items by injecting  and fusing high-order semantics. The proposed HGRec first adopts embedding layer to initialize node embedding. Then, semantic aggregation layer and semantic fusion layer will inject high-order semantic into node embedding via multi-hops meta-path and fuses rich semantics via multiple meta-paths, respectively. Lastly, we  leverage the fused embedding of user and item for recommendation.
\subsection{Embedding Initialization}
Following the previous works \cite{he2017neural,wang2019neural}, we random initialize node embedding matrix and use look-up to get the initial embedding of  user $u$ and item $i$, denoted as $\mathbf{e}_u\in \mathbb{R}^d$ and  $\mathbf{e}_i \in \mathbb{R}^d $, respectively. Here $d$ is the dimension of node embedding.

\subsection{Semantic Aggregation  Layer}
After obtain initial node embedding, we  propose semantic aggregating layer to aggregate multi-hops meta-path based neighbors and update node embedding, so the high-order semantic information is well preserved. For clearly,  we first introduce the first-order aggregation in semantic aggregation layer and then generalize it to multiple
successive layers (aka. high-order semantic aggregation).

\textbf{First-order Semantic Aggregation}
Taking one user $u$ and one user-related meta-path $\Phi^U$ as an example, 
we propose semantic aggregation layer $\mathcal{A}$  to aggregate  meta-path based neighbors $\mathcal{N}^{\Phi^U}_u$ and get the first-order user embedding $\mathbf{e}_{u }^{\Phi^U, 1}$, shown as follows:
\begin{equation}
    \mathbf{e}_{u }^{\Phi^U, 1}=\mathcal{A}(u, \Phi^U).
\end{equation}
Rather than simple neighbor combination, we consider the complex interaction between node and its neighbors in  aggregating process.  Specifically, we encode the interaction between node $u$ and its neighbor $k$  into aggregating process via $\mathbf{e}_{k} \odot \mathbf{e}_{u}$, where $\odot$ denotes the element-wise product. The overall aggregating process is shown as follows:

\begin{equation}
\mathbf{e}_{u }^{\Phi^U, 1}= \mathbf{W}_{1}^{\Phi^U} \mathbf{e}_{u}  + \sum_{k\in \mathcal{N}^{\Phi^U}_u}\left(\mathbf{W}_{1}^{\Phi^U} \mathbf{e}_{k}+\mathbf{W}_{2}^{\Phi^U}\left(\mathbf{e}_{k} \odot \mathbf{e}_{u}\right)\right) ,
\end{equation}
where 
$\mathbf{W}_{1}^{\Phi^U},\mathbf{W}_{2}^{\Phi^U}$ are weight matrixes.
The first-order semantic aggregation only aggregates one-hop meta-path based neighbors  into node embedding, while high-order semantic revealed by multi-hops meta-path plays a crucial role in improving node embedding.

\textbf{High-order Semantic Aggregation} Considering the high-order semantic revealed by multi-hops meta-path, we  stack first-order  semantic aggregation for multiple layers and recurrently aggregate  corresponding meta-path based neighbors, 
so the  high-order semantic is injected into node embedding, shown as follows:
\begin{equation}
\mathbf{e}_{u }^{\Phi^U, L}=\mathcal{A}^L(\cdots \mathcal{A}^2(\mathcal{A}^1(u, \Phi^U))),
\end{equation}
where $\mathbf{e}_{u }^{\Phi^U, L}$ denotes the $L$-order user embedding. 
Then, we concatenate different order user embedding and get the semantic-specific embedding of user $u$, shown as follows:
\begin{equation}
    \mathbf{e}_{u }^{\Phi^U}=\mathbf{e}_{u }^{\Phi^U, 1}||\mathbf{e}_{u }^{\Phi^U, 2}||,\cdots, ||\mathbf{e}_{u }^{\Phi^U, L},
\end{equation}
where $||$ is the concatenation operation. However, one meta-path cannot comprehensively describe the characteristics of node from different aspects.  Considering a set of user-related meta-paths $\{ \Phi_1^U, \Phi_2^U, \cdots, \Phi_{K_1}^U\}$, we can get $K_1$ groups of user embeddings $ \{\mathbf{E}_{u }^{\Phi_1^U},\mathbf{E}_{u }^{\Phi_2^U},\cdots,\mathbf{E}_{u }^{\Phi_{K_1}^U} \}$.

Similar to user embedding, given a set of item-related meta-paths $\{ \Phi_1^I, \Phi_2^I, \cdots, \Phi_{K_2}^I\}$, we can get $K_2$ groups of item embeddings $\{\mathbf{E}_{i }^{\Phi_1^I},\mathbf{E}_{i }^{\Phi_2^I},\cdots,\mathbf{E}_{i }^{\Phi_{K_2}^I} \}$.

\subsection{Semantic Fusion Layer}
After obtaining multiple higher-order node embedding, we need to learn the importance of different meta-paths and fuse them properly for better recommendation. Given $K_1$ groups of user embeddings $ \{\mathbf{E}_{u }^{\Phi_1^U},\mathbf{E}_{u }^{\Phi_2^U},\cdots,\mathbf{E}_{u }^{\Phi_{K_1}^U} \}$, we propose semantic fusion layer $\mathcal{F}$ to learn the weights of different meta-paths (e.g., $w^{\Phi_1^U}, w^{\Phi_2^U}, \cdots, w^{\Phi_{K_1}^U}$), shown as follows:

\begin{equation}
    (w^{\Phi_1^U}, w^{\Phi_2^U}, \cdots, w^{\Phi_{K_1}^U}) = \mathcal{F}(\mathbf{E}_{u }^{\Phi_1^U},\mathbf{E}_{u }^{\Phi_2^U},\cdots,\mathbf{E}_{u }^{\Phi_{K_1}^U}).
\end{equation}
To learn the  importance of each meta-path (e.g., $\alpha_{\Phi _{k}^U}$), we first project node embedding into the  attention space and then use a semantic attention vector  $\mathbf{q}_{U}$ to measure the importance  of  meta-path specific embedding, 
\begin{equation}
    \alpha_{\Phi _{k}^U}=\frac{1}{\left |V \right |}\sum_{i\in V }\mathbf{q}_U^{\top}\cdot \tanh\left (\mathbf{W}_U\cdot \mathbf{e}_{u}^{\Phi^U_k }+\mathbf{b}_U\right ),
\end{equation}
where $\mathbf{W}_U$ and $\mathbf{b}_U$ are weight and bias, respectively.
Then, we normalize them via softmax function and  get meta-path weights $w_{\Phi _{k}^U}$, shown as follows:
\begin{equation}
   w_{\Phi _{k}^U}=\frac{\exp\left ( w_{\Phi _{k}^U} \right )}{\sum_{k=1}^{K_1}\exp\left ( w_{\Phi _{k}^U} \right )}.
\end{equation}
With the learned weights as coefficients, we
can fuse  multiple user embeddings to obtain the final
embedding $\mathbf{E}_u$ as follows:

\begin{equation}
    \mathbf{E}_u=\sum_{k=1}^{K_1} w_{\Phi _{k}^U}\cdot \mathbf{E}_u^{\Phi _{k}^U}.
\end{equation}

Similar to user embedding,   we can fuse $K_2$ groups of item embeddings $\{\mathbf{E}_{i }^{\Phi_1^U},\mathbf{E}_{i }^{\Phi_2^U},\cdots,\mathbf{E}_{i }^{\Phi_{K_2}^U} \}$ and  obtain the final
embedding  of item $\mathbf{E}_i$.

\subsection{Model Prediction}
The final part of the model is to recommend items for users based on their embedding. Here  we calculate the inner product  of user and item for recommendation, as follows:

\begin{equation}
    \hat{y}_{ui}=(\mathbf{E}_{u })^\top \mathbf{E}_{i}.
\end{equation}
Then, we calculate BPR loss \cite{wang2019neural} and optimize the parameters, as follows:
\begin{equation}
    L=\sum_{\left ( u,i,j \right )\in \mathcal{O}}-\ln\sigma \left ( \hat{y}_{ui}- \hat{y}_{uj} \right )+\lambda \left \| \Theta  \right \|_{2}^{2},
\end{equation}
where $\mathcal{O}=\left\{(u, i, j) |(u, i) \in \mathcal{R}^{+},(u, j) \in \mathcal{R}^{-}\right\}$ denotes the pairwise
training data,$ \mathcal{R}^{+}$ indicates the observed interactions,  $\mathcal{R}^{-}$ is
the unobserved interactions, $\Theta $ denotes all trainable model parameters, and $\lambda $
controls the L2 regularization strength to prevent overfitting.

\section{EXPERIMENTS}


We conduct  experiments on three heterogeneous graphs: Amazon, Yelp and Movielens (details are shown in Table \ref{dataset}).
\begin{table}[]
\caption{Statistics of the datasets.} 
\label{dataset}
\begin{tabular}{|l|l|l|l|l|}
\hline
Datasets                   & Relation(A-B)     & \#A   & \#B   & \#A-B   \\ \hline
\multirow{3}{*}{Movielens} & User - Movie    & 943   & 1,682 & 100,000 \\ \cline{2-5} 
                           & User - User    & 943   & 943   & 47,150  \\ \cline{2-5} 
                           & Movie - Genre     & 1,682 & 18    & 2,861   \\ \hline
\multirow{4}{*}{Amazon}    & User - Item     & 6,170 & 2,753 & 195,791 \\ \cline{2-5} 
                           & Item - Cate.   & 2,753 & 22    & 5,508   \\ \cline{2-5} 
                           & Item - Brand      & 2,753 & 334   & 2,753   \\ \hline
\multirow{4}{*}{Yelp}    & User - Item     & 16,239 & 14,284 & 198,397 \\ \cline{2-5} 
                           &User - User    & 16,239   & 16,239   & 158,590  \\ \cline{2-5} 
                           & Item - City      & 14,284 & 47   & 14,267   \\ \hline                        
\end{tabular}
\end{table}
We  compare with some state-of-art baselines, include BPRMF \cite{bpr}, NMF \cite{he2017neural}, GAT \cite{18iclr_gat}, MCRec \cite{hu2018leveraging}, NGCF \cite{wang2019neural}, to verify the effectiveness of the proposed model. We also test a variant of HGRec, denotes as HGRec-, which assigns the same importance to each meta-path.

For evaluation, we split  datasets into training set and test set with 8:2 ratio and employ Pre@10, Recall@10, HR@10 and NDCG@10 as evaluation metrics. 

We randomly initialize parameters and optimize models with Adam. For the proposed HGRec, we set the L2 regularization  to 1e-2, the dimension of the semantic attention vector $\mathbf{q}$ to 64,  the dropout to 0.8,  and 
the learning rate to 5e-4, 1e-3 and 5e-3 on Movielens  Amazon and  Yelp, respectively.
We also use early stopping with a patience of 100 to aviod overfitting.

\subsection{Overall Performance Analysis}
The experiment results   are shown in Table \ref{perform} and  we have the following   observations:.


\begin{table}[]
\caption{Overall Performance Comparison }
\label{perform}
\begin{tabular}{|l|c|c|c|c|}
\hline
\multirow{2}{*}{Models} & \multicolumn{4}{c|}{Movielens}              \\ \cline{2-5} 
                       & Pre@10  & Rec@10 & NDCG@10 & HR@10 \\ \hline
BMF                    & 0.3251 & 0.2096   & 0.4081 & 0.8928     \\ \hline
NMF                    & 0.1704 & 0.1163   & 0.2336 & 0.7739     \\ \hline
GAT                    & 0.2068 & 0.1210   & 0.2556  & 0.7548     \\ \hline
MCRec                  & 0.3310  & 0.2129    & 0.2624  & 0.9025      \\ \hline
NGCF                   & 0.3369 & 0.2179   & 0.4178 & 0.9045     \\ \hline
HGRec-            & \textbf{0.3670} & \textbf{0.2412}   & \textbf{0.4551} & 0.9172     \\ \hline
HGRec                  & 0.3667 & 0.2405   & 0.4547 & \textbf{0.9193}    \\ \hline
Improv.                    & 6.70$\%$ &6.80$\%$   & 8.19$\%$ & 1.36$\%$  \\ \hline
\end{tabular}
\end{table}

\begin{table}[]
\begin{tabular}{|l|c|c|c|c|}
\hline
\multirow{2}{*}{Models} & \multicolumn{4}{c|}{Amazon}               \\ \cline{2-5} 
                       & Pre@10  & Rec@10 & NDCG@10 & HR@10 \\ \hline
BMF                    & 0.0490 & 0.0881   & 0.1176 & 0.3232     \\ \hline
NMF                    & 0.0168 & 0.0264   & 0.0463 & 0.1371     \\ \hline
GAT                    & 0.0410 & 0.0810  & 0.1096  & 0.2998     \\ \hline
MCRec                    & 0.0309 & 0.0697  & 0.1131  & 0.3027     \\ \hline
NGCF                   & 0.0495 & 0.0870  & 0.1150 &0.3224     \\ \hline
HGRec-                 & 0.0553 & 0.0988   & 0.1313 & 0.3503      \\ \hline
HGRec                  & \textbf{0.0588} & \textbf{0.1054}   & \textbf{0.1384} & \textbf{0.3746}   \\ \hline
Improv.                   &5.95$\%$ &6.26$\%$   &5.13$\%$  & 6.48$\%$   \\ \hline
\end{tabular}
\end{table}

\begin{table}[]
\begin{tabular}{|l|c|c|c|c|}
\hline
\multirow{2}{*}{Models} & \multicolumn{4}{c|}{Yelp}               \\ \cline{2-5} 
                       & Pre@10  & Rec@10 & NDCG@10 & HR@10 \\ \hline
BMF                    & 0.0039 & 0.0287   & 0.0150 & 0.0291     \\ \hline
NMF                    &0.0012 & 0.0265   & 0.0233 & 0.0398     \\ \hline
GAT                    &0.0038  & 0.0240  & 0.0171  & 0.0363     \\ \hline
MCRec                    & 0.0031 & 0.0531  & 0.0201  & 0.0432     \\ \hline
NGCF                   & 0.0073 & 0.0410  & 0.0271 &0.0667     \\ \hline
HGRec-                 & 0.0076 & 0.0433   & 0.0237 & 0.0506      \\ \hline
HGRec                  & \textbf{0.0078} & \textbf{0.0447}   & \textbf{0.0310} & \textbf{0.0671}   \\ \hline
Improv.                   &6.41$\%$ &3.13$\%$   &12.6$\%$  &1.03$\%$   \\ \hline
\end{tabular}
\end{table}

\textbullet ~The proposed  HGRec consistently performances  better than  baselines with significant gap on all the datasets. In particular, HGRec improves over the strongest baseline NGCF  w.r.t.
Recall@10 by 6.80\%, 17.49\%,3.13\%  in Movielens
Amazon and Yelp, respectively. The results demonstrate that   injecting   rich high-order semantics  into the node embedding  indeed improves the recommendation performance.

\textbullet ~Compare HGRec with HGRec- , we can observe that HGRec outperforms HGRec-  on all  datasets. This proves that the semantic fusion layer is able to identify the importance of meta-paths and then  enhance the performance of HGRec.

\textbullet ~Graph neural network based recommendation models show their superiorities over traditional MF based models,  demonstrating the importance of nonlinear structural
interactions among nodes.
\begin{table}[]
\caption{Effectiveness of Layer Number}
\label{movie_depth}
\begin{tabular}{|l|c|c|c|c|}
\hline
\multirow{2}{*}{L} & \multicolumn{2}{c|}{Movielens} & \multicolumn{2}{c|}{Amazon} \\\cline{2-5} 
                  & Rec@10      & NDCG@10      & Rec@10     & NDCG@10   \\\hline
1           & 0.2390        & 0.4506      & 0.0947       & 0.1251    \\\hline
2           & 0.2391        & 0.4526        & 0.0864       & 0.1151    \\\hline
3           & \textbf{0.2405}        & \textbf{0.4547}      & \textbf{0.1054}       & \textbf{0.1384}    \\\hline
 4          & 0.2391        & 0.4513      & 0.0743       & 0.1064  \\ \hline
\end{tabular}
\end{table}
\useunder{\uline}{\ul}{}

\textbf{Effect of Layer Numbers.} 
To investigate  the whether high-order semantic  improves node embedding, we vary the model depth (e.g., $L=1,2,3,4$) and show the results on Table \ref{movie_depth}. We can find that with the growth of model depth, the performance of HGRec  are sustainable growth and achieves the best performance
when   $L$ is set to 3,  indicating the effectiveness of high-order semantic. After that, the performance of
HGRec starts to degenerate which may because of overfitting.

\section{Conclusion and Future Work}
In this work, we highlight the critical importance of rich  high-order semantics in improving node embedding for better recommendation. Specifically, we design a semantic aggregation layer which aggregates multi-hop meta-path neighbors so as to inject high-order semantic into node embedding.  To describe the characteristics of node comprehensively, we leverage a semantic fusion layer to  fuse rich semantic revealed by multiple meta-paths. Experimental results demonstrates the superiority of the proposed model and show the potentially good interpretability for the recommendation results.




\nocite{langley00}

\bibliography{ref}
\bibliographystyle{icml2020}





\end{document}